# Probability density function and detection threshold in high contrast imaging with partially polarized light

Natalia Yaitskova

European Organization for Astronomical Research in the Southern Hemisphere (ESO),

Karl Schwarzschildstr. 2, D-85748 Garching b. Muenchen, Germany

We obtain an expression for the probability density function (PDF) of partially developed speckles formed by light with an arbitrary degree of polarization. From the probability density we calculate the detection threshold corresponding to the $5\sigma$ confidence level of a normal distribution. We show that unpolarized light has an advantage in high contrast imaging for low ratios of the deterministic part of the point spread function (DL PSF) to the halo, typical in coronagraphy. © 2009 Optical Society of America

The speckle phenomena in laser optics created a great interest about 40 years ago. Now this interest grows gain in the new context of high contrast astronomical imaging. Imaging and characterization of extra-solar planets close to their parent star is extremely difficult, because it involves detection at very low signal to noise ratios. One of the main contributors to the noise is quasi-static speckles of the stellar light, coursed by the imperfections in the imaging optics. The knowledge of the statistical properties of these speckles is essential for developing smart detection algorithms. In this Letter we do not address the detection itself, but the statistics of stellar speckles. A number of authors have used a modified Rician distribution for the PDF of stellar speckles [1-3]. This distribution was obtained assuming *fully polarized* light [4,5], so the results cannot be applied directly to the case of astronomical images. The light from a star is known to be *unpolarized* [6]. Imaging optics introduces some polarization [7], so in general the speckles from the star arriving to the detector are *partially polarized*.

Below we calculate the PDF for the intensity of speckles appearing on the constant background of the DL PSF for the case of an arbitrary degree of polarization. These are the pinned speckles and the speckles of the second order. The instantaneous intensity of partially polarized light with an arbitrary degree of polarization $P$ can be expressed as a sum of two statistically independent undeveloped speckle patterns [4]:

$$I = I_1 + I_2, \tag{1}$$

Each of the intensity patterns is a sum of squares of real and imaginary parts of the complex amplitude, for which we use a classical assumption of a non-central Gaussian distribution. $I_1$ and $I_2$ follow a non-central chi-square distribution of the second order [4]. For example, for the first component the PDF is:

$$p(I_1) = \frac{1}{H_1} \exp\left(-\frac{I_1 + D_1}{H_1}\right) I_0\left(2\frac{\sqrt{I_1 D_1}}{H_1}\right), \tag{2}$$

where $I_0$ is a modified Bessel function of zero order. Parameters of the distribution, $H$ and $D$, depend on the degree of polarization as:

$$D_{1(2)} = \frac{1}{2}(1 \pm P)|C(w)|^2, \quad H_{1(2)} = \frac{1}{2}(1 \pm P)\langle|S(w)|^2\rangle, \tag{3}$$

where *plus* is taken for the speckle pattern $I_1$ and *minus* is taken for the speckle pattern $I_2$. Eq.(2) is also known as a modified Rician distribution. The sum $D_1 + D_2 = |C(w)|^2$ is the DL PSF, the deterministic part of the PSF defined by the optical system without random disturbances. It can be also a PSF reduced to some extend by a coronagraph. The sum $H_1 + H_2 = \langle|S(w)|^2\rangle$ is the halo, an ensemble average speckled field on top of the DL PSF [8]. Both functions depend on the coordinate on the image plane $w$. The mean level of the total intensity equals to the DL PSF plus the halo: $\langle I \rangle = |C(w)|^2 + \langle|S(w)|^2\rangle$. We write the PDF for the normalized intensities:

$$p_{1(2)}(x_{1(2)}) = 2\frac{1+r}{1 \pm P}\exp\left[-\frac{2x_{1(2)}(1+r)}{1 \pm P} - r\right] I_0\left[2\sqrt{\frac{2x_{1(2)}(1+r)r}{1 \pm P}}\right], \tag{4}$$

where $x_{1(2)} = I_{1(2)}/\langle I \rangle$ is a normalized intensity for the first (second) speckle pattern, and $r$ is the ratio of the DL PSF to the halo, $r = |C(w)|^2/\langle|S(w)|^2\rangle$. For an ideal coronagraph entirely removing the deterministic part of the PSF $r = 0$. The PDF of the total field can be calculated either by numerical convolution of $p_1(x_1)$ and $p_2(x_2)$ [9], or by the method of the characteristic function [10]. The characteristic functions for the PDFs from Eq. (4) are

$$\Xi_{1(2)}(z) = \exp\left[\frac{r}{1-iz\frac{1\pm P}{2(1+r)}} - r\right] \frac{1}{1-iz\frac{1\pm P}{2(1+r)}}, \tag{5}$$

where "plus" is taken for $\Xi_1$ and "minus" for $\Xi_2$. The PDF of the total normalized intensity, $x = I/\langle I \rangle$, is the inverse Fourier transform of the product $\Xi_1(z)\Xi_2(z)$. To proceed with the calculations we expand each of the exponential functions in Eq.(5) in a sum and use the following table integral:

$$\frac{1}{2\pi}\int_{-\infty}^{\infty}(\alpha-iz)^{-(l+1)}(\beta-iz)^{-(k+1)}\exp(-izx)dz$$
$$= \exp(-\alpha x)\frac{x^{k+l+1}}{\Gamma(k+l+2)}M[k+1,k+l+2,(\alpha-\beta)x]. \tag{6}$$

where $M[.,.,.]$ is the degenerated hypergeometric function, also known as Kummer function, $\Gamma(...)$ is the Gamma function [11]. After re-combination of terms the final expression for the PDF becomes:

$$p(x) = 4x\frac{(1+r)^2}{1-P^2}\exp\left\{-\frac{2}{1-P}[x(1+r)+r(1-P)]\right\}$$
$$\sum_{l,k=0}^{\infty}\frac{1}{\Gamma(k+l+2)k!l!}\left[\frac{2r(1+r)x}{1-P}\right]^l\left[\frac{2r(1+r)x}{1+P}\right]^k M\left[k+1,k+l+2,\frac{4xP(1+r)}{1-P^2}\right]. \tag{7}$$

To our knowledge, the PDF in this form for an arbitrary combination of $P$ and $r$ is unpublished. A limiting case $P=1$ leaves only $I_1$ in Eq.(1), so Eq. (7) merges with the non-central chi-square distribution of the second order given by Eq.(2) with $P=1$ and $x_1 = x$. To prove it we use the following limit for Kummer function:

$M[a,b,z]\underset{z\to\infty}{=}\frac{\Gamma(b)}{\Gamma(a)}e^z z^{a-b}$. Substituting it into Eq.(7) allows factorizing the two

sums, one of which gives a Bessel function of zero order and the other an exponential function. For $P=0$ Eq. (1) is a sum of squares of four identically distributed normal values, hence the distribution in this case must coincide with the chi-square distribution of the fourth order. The proof is based on the fact that for any $a$ and $b$: $M[a,b,0]=1$. Afterwards we apply a multiplication theorem for the Bessel functions [11]. The PDF in this case is [10]:

$$p(x,P=0)=4(1+r)^2 2x\exp[-2(1+r)x-2r]\frac{I_1\left[4\sqrt{r(1+r)x}\right]}{4\sqrt{r(1+r)x}}, \qquad (8)$$

where $I_1$ is a modified Bessel function of degree one. Analogously we demonstrate that for $r=0$ Eq.(7) merges with the PDF for fully developed partially polarized speckles.

The PDF is shown in Figure 1 on example when the DL PSF is equal to the halo ($r=1$). When the degree of polarization decreases the distribution becomes more localized around the mean, which indicates that the fluctuation of speckles decreases. It comes from the representation of partially polarized light as the sum of two independent speckle patterns. If two or more speckle patterns are added up, the total field is smoothed out, and the intensity fluctuation is lower. The difference is essential for the points in the image plane where $r\leq 1$. This situation is met in coronagraphic imaging. For observations without a coronagraph $r\gg 1$ in all areas interesting for the detection and the gain is not significant. In this regime, the distribution looks similar to a Gaussian.

Provided the knowledge about the PDF we now calculate the detection threshold corresponding to a given confidence level. For Gaussian statistics, the conventional 5σ level answers to a $CL_{5\sigma} = 0.9999997125$ confidence level. It means that if the statistics of speckle intensity were Gaussian, a probability of finding a speckle with an intensity higher than $I_{th} = \langle I \rangle + 5\sigma_I$ would be equal to $1 - CL_{5\sigma} = 2.875 \cdot 10^{-7}$. The application of the 5σ rule to the probability density found in Eq. (7) leads to a lower confidence level. We now wish to find a detection threshold $n(P,r)$ for this PDF providing the same confidence level. For the normalized intensity it means: $prob[x > 1 + n(P,r)\sigma_x(P,r)] = 1 - CL_{5\sigma}$. The standard deviation $\sigma_x(P,r)$ can be calculated exactly as a square root of a sum of two variances for $x_1$ and $x_2$:

$$\sigma_x(P,r) = \sqrt{\frac{1+P^2}{2}} \frac{\sqrt{1+2r}}{1+r} \qquad (9)$$

The function $n(P,r)$, calculated from Eq.(7), is shown in Figure 2. When $r \ll 1$, a threshold $n=14$ must be considered for fully polarized light and $n=11$ for unpolarized light. When $r \sim 1$ there is still a difference in threshold: $n=11$ for $P=1$ and $n=9$ for $P=0$. Therefore, for small and medium values of $r$ *a lower threshold is required to reach the same confidence level of detection with unpolarized light, compared to polarized light.* For large $r$ the threshold saturates to the level $n=5$ corresponding to the normal distribution.

For the high contrast imaging (HCI) the value of $r$ reflects the performance of a coronagraph meant to reduce the DL PSF with respect to the performance of extreme adaptive optics (XAO) meant to decrease the halo. Small $r$ (at a given angular distance) means that the coronagraph outperforms the XAO; large $r$ means the opposite. To help designing a HCI instrument for any predicted performance of coronagraph and XAO we find an analytical expression approximating the curves from Figure 2. It estimates the detection threshold for arbitrary $r$ and arbitrary degree of polarization required to achieve a confidence level equivalent to the 5$\sigma$ level of Gaussian distribution:

$$n(r,P) \approx [n(0,P)-5]\frac{\sqrt{1+2.4\cdot 2r}}{1+2.4r}+5. \qquad (10)$$

Here $n(0,P)$ is the threshold in the case of fully developed speckles (Figure 3).

**List of Figures**

**Figure 1**. Probability density functions for different values of the degree of polarization when the DL PSF is equal to the halo

**Figure 2**. Detection threshold as a function of the ratio between the DL PSF and the halo. Curves are shown for three different values of the degrees of polarization

**Figure 3**. Detection threshold as a function of the degree of polarization in absence of the DL PSF

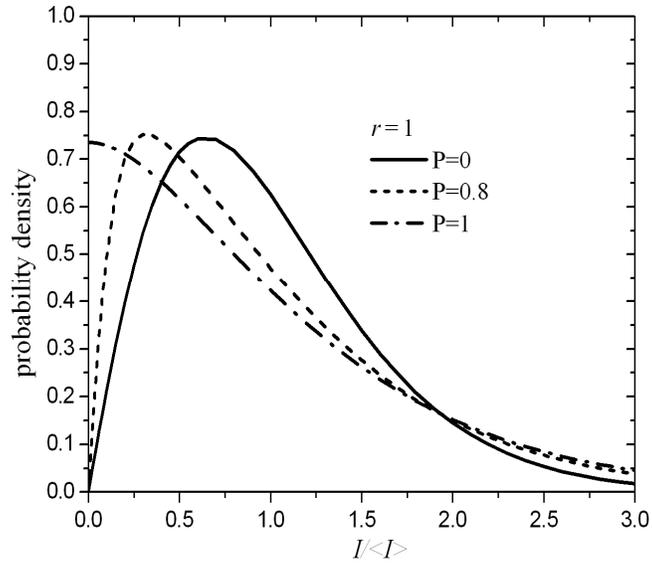

Figure 1. Probability density functions for different values of the degree of polarization when the DL PSF is equal to the halo

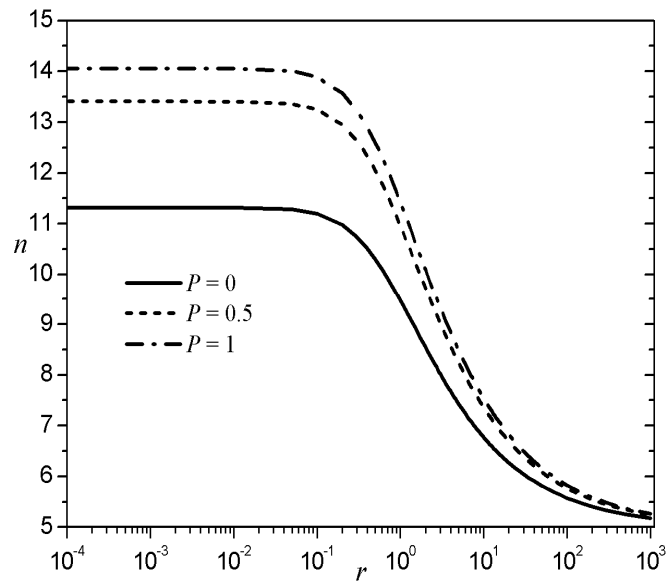

Figure 2. Detection threshold as a function of the ratio between the DL PSF and the halo. Curves are shown for three different values of the degrees of polarization

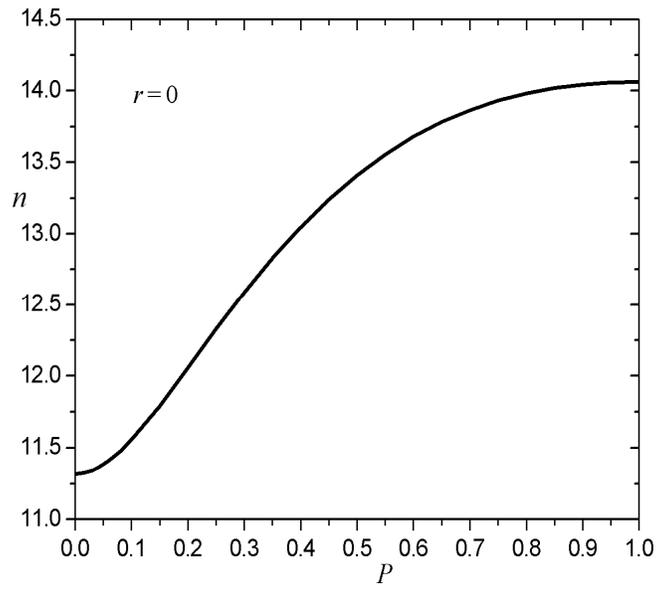

Figure 3. Detection threshold as a function of the degree of polarization in absence of the DL PSF